# Annihilation-limited Long-range Exciton Transport in High-mobility Conjugated Copolymer Films

Yuping Shi[a,b,c,*], Partha P. Roy[a,b,d], Naoki Higashitarumizu[e,f,g], Tsung-Yen Lee[a], Quanwei Li[a], Ali Javey[e,f], Katharina Landfester[c], Iain McCulloch[h,i], and Graham R. Fleming[a,b,*]

[a]Department of Chemistry, University of California, Berkeley, CA 94720, USA.

[b]Molecular Biophysics and Integrated Bioimaging Division, Lawrence Berkeley National Laboratory, Berkeley, CA 94720, USA.

[c]Max Planck Institute for Polymer Research, Mainz 55128, Germany.

[d]Department of Chemistry, Northwestern University, IL 60208, USA.

[e]Electrical Engineering and Computer Sciences, University of California, Berkeley, CA 94720, USA.

[f]Materials Sciences Division, Lawrence Berkeley National Laboratory, Berkeley, CA 94720, USA.

[g]JST, PRESTO, 4-1-8 Honcho, Kawaguchi, Saitama 332-0012, Japan.

[h]Department of Chemistry, University of Oxford, Oxford OX1 3TA, UK.

[i]Department of Electrical and Computer Engineering, Princeton University, New Jersey 08544, USA.

[*]Corresponding Authors. Email: shiy@mpip-mainz.mpg.de (Y.S.); grfleming@lbl.gov (G.R.F)

## Significance Statement

The practical applications of solution-processable semiconductor polymers are often limited by the speed and distance of travel of exciton motion along with the excitation loss during transport. Here we use phase-cycled transient absorption spectroscopy, quantitatively separating the single-, two-, and three-particle nonlinear responses in high-mobility planar conjugated copolymers. We find interchain transport in device-relevant continuous thin films with exciton diffusion lengths exceeding the film thickness, along with an exciton-exciton annihilation limited transport regime from room temperature down to 77 K, rather than the classical diffusion-limited annihilation. Remarkably, we find that excitation loss from many-body interactions is suppressed in the thin film as compared to isolated polymer chains in solution.

# Abstract

A combination of ultrafast, long-range and low-loss excitation energy transfer from the photo-receptor location to a functionally active site is essential for cost-effective polymeric semiconductors (1–4). Delocalized electronic wavefunctions along π-conjugated polymer backbone can enable efficient intrachain transport (5, 6), while interchain transport is generally thought slow and lossy due to weak chain-chain interactions (7, 8). In contrast to the conventional strategy of mitigating structural disorder (9–11), amorphous layers of rigid conjugated polymers, exemplified by highly planar poly(indacenodithiophene-co-benzothiadiazole) (IDT-BT) donor-acceptor copolymer, exhibit trap-free transistor performance and charge-carrier mobilities similar to amorphous silicon (12–15). Here we report long-range exciton transport in *HJ*-aggregated IDTBT thin-film, in which the competing exciton transport and exciton-exciton annihilation (EEA) dynamics are spectroscopically separated using a phase-cycling-based scheme (16) and shown to depart from the classical diffusion-limited and strong-coupling regime. In the thin film, we find an annihilation-limited mechanism with ≪100% per-encounter annihilation probability, facilitating the minimization of EEA-induced excitation losses. In contrast, excitons on isolated IDTBT chains diffuse over 350 nm with 0.56 cm$^2$ s$^{-1}$ diffusivity, before eventually annihilating with unit probability on first contact. We complement the pump-probe studies with temperature dependent photocurrent and EEA measurements from 295 K to 77 K and find a remarkable correspondence of annihilation rate and photocurrent activation energies in the 140 K to 295 K temperature range.

Solution-processable polymeric semiconductors are finding applications requiring excellent performance in optoelectronics (17–20), photoelectrocatalysis (21–23), artificial photosynthesis (24) and biomedicine (25–27) due to their tunable energy levels, the soft morphology and scalable production. Continuing improvement in effective carrier mobility and diffusion length in polymer-based devices has been accomplished by making use of long-range structural order and relatively strong on-chain electronic coupling (9, 28). In "conventional" semiconducting polymers, achievement of sufficient local structural ordering, e.g., in terms of fine-tuned chain conformation and interchain packing, often requires additional chain orientation techniques in order to tailor the delocalization length and local strength of electronic interactions and excitation transport potential for intrachain and interchain transport (10). In contrast to this conventional strategy of creating well-ordered microstructures (9–11), a combination of adaptive rigid planar backbone conformations with nearly amorphous chain-chain interconnection in continuous layers of donor-acceptor copolymers, exemplified by large-area extended poly(indacenodithiophene-co-benzothiadiazole) (IDT-BT) thin films resulting directly from spin-coating fabrication, has been found to enable exceptionally high charge-carrier mobility and nearly trap-free transport properties (12–15).

Under real device operating conditions, the prevalent structural disorder and concomitant energetic dispersion ($\sigma$) in an active conjugated polymer deposit, resulting from spatially varying chain conformations and local traps, dictates the overall transport performance. Power efficiency and other performance is generally assumed to be detrimentally affected by interchain interactions and hopping (29, 30). In contrast, in this study we find that the presence of interchain transport in large-area extended amorphous films of the high-mobility IDTBT copolymer suppresses excitation loss associated with many-body interactions and exciton-exciton annihilation (EEA), compared with well-isolated IDTBT chains in solution.

In such systems it is nearly impossible to record high time resolution dynamics with acceptable signal to noise ratio without placing multiple excitations within range of each other. Interactions on encounter of two excitons (Fig. 1a – b) lead to exciton-exciton annihilation (EEA) (Fig. 1c) and loss of one of the two excitons. Such a loss competes with transport and becomes dominant at high excitation densities required in a wide range of devices. On the other hand, analysis of EEA dynamics holds promise of providing a fuller understanding of the EEA for improved material and device design.

The presence of multiple excitations in the same system means that the optical response cannot be described as a third order response and instead also contains contributions from the fifth order, seventh order, etc., responses. In much previous work the fact that the signal contained multiple superimposed contributions was modelled at the simplest level by assuming that the higher order nonlinear responses could be described by a term proportional to $n(t)^2$ with $n$ the exciton density so that $dn(t)/dt = -k_e n(t) + \gamma n(t)^2$. This phenomenological expression, however, does not give microscopic insight into the response.

The form of the 5$^{th}$ order and higher responses clearly depends on the underlying microscopic dynamics. In the case that exciton numbers dominate the response Malý et al (16) derived explicit expressions for the 5$^{th}$ and 7$^{th}$ order responses that provide microscopic insight into the nature of the exciton motion and the probability of annihilation on encounter (always assumed to be 100% previously). The conventional picture considers excitation transport dynamics in semiconducting polymer structures to be incoherent and governed by a diffusion-limited mechanism (31, 32). Within this classical transport framework, the non-radiative EEA is assumed to occur extremely rapidly with a 100% probability as a result of strong electronic coupling when the two excitons encounter each other within an annihilation radius.

In the case of IDTBT thin-films, we instead find an annihilation-limited transport mechanism from 295 K to 77 K, in which excitons are allowed to encounter and separate several times before they eventually annihilate within the microstructural registry and transport network in the solid state. Our analysis gives diffusion constants that are among the highest reported in continuous device-relevant films of polymeric semiconductors. We also find that interchain hopping is thermally activated and is the rate limiting step in both EEA and photocurrent generation in IDTBT thin-films, pinpointing a similar structural and mechanistic base for the two essential processes.

## Experimental Results

**Electronic Spectra and Aggregation State**

The IDTBT copolymer is designed and synthesized with a rigid ring-fused IDT donor moiety, in which the two symmetric $sp^3$-hybridized bridging carbon atoms and a high potential barrier to torsion arising from the electrostatic attraction between the IDT peripheral hydrogen and the BT nitrogen atom result into a remarkably planar backbone conformation (12, 15, 32), in spite of the steric hindrance effect of the $sp^3$-carbon atoms promoting a nearly amorphous chain-chain microstructure in films of the copolymer. The highest occupied molecular orbital (HOMO) of IDTBT is delocalized over the whole chain but the lowest unoccupied molecular orbital (LUMO) is mostly localized on the BT units (12); the spatial overlap between HOMO and LUMO orbitals on an IDTBT backbone induces a strong on-chain charge transfer character, enabling ultrafast excitation migration along the backbone and efficient decays of IDTBT excited states. The optical absorption spectrum of an IDTBT thin film shows two main electronic transitions in the visible spectral range, with the red band exhibiting two clear peaks which correspond to the 0-0 and 0-1 vibronic transitions (Fig. 1d). The spectrum is sharper in thin films than in solution due to a decrease in inhomogeneous broadening and/or the stabilization of more extended chain conformations in the solid state.

In order to analyze the IDTBT aggregation state and photophysical behavior, we recorded temperature-dependent linear absorption and emission spectra in pristine IDTBT thin-films (Fig. 1d-e). As the temperature decreases, the UV-vis and emission spectra show a gradual red-shift and narrowing, and maintain a small and nearly constant Stokes shift (42 ± 2 nm = 0.1 eV). These findings imply an extremely high degree of conformational rigidity and linearity of the IDTBT backbone. In addition, an increase in emission intensity and increase in the 0-0 to 0-1 peak ratio in both absorption and emission spectra are noticeable. Typically, the temperature dependence of the 0-0 to 0-1 emission peak ratio in a molecular aggregate depends on the relative orientation of the transition dipole moments (*J* vs *H*-type) (33, 34). Isolated single polymer chains behave as *J*-aggregates due to the head-to-tail arrangement and strong intermolecular coupling of alternating donor-acceptor monomeric unit (IDT-BT), which is evident in the long fluorescence lifetime in solution (1.59 ns shown in Fig. S1). On the other hand, the interchain short-contacts in a thin-film can lead to *H*-type of interaction resulting in weakly emissive or optically dark interchain excitonic states. The 0-0 to 0-1 emission peak ratio in an IDTBT polymer thin film exhibits a $T^{-1}\exp(-\Delta/k_BT)$ dependence (33) for the measured temperature range (with an estimate of a narrow interchain exciton bandwidth $\Delta = 23.6 \pm 3.5$ cm$^{-1} \approx 3$ meV in Fig. 1f and Fig. S2; an estimate of such low $\Delta$ implies the formation of extremely straight backbone conformations and/or weak interchain coupling at chain-chain crossing points). This contrasts with the behavior of *J*-aggregates which show a $T^{-1/2}$ dependence (34), and instead follows the typical temperature-dependent spectral signature of a *HJ*-aggregate as predicted by Spano and coworkers (33). This assignment is further supported by the substantial enhancement of the emission yield at lower temperature: the 0-0 emission peak intensity at 77 K is ~10 times greater than that of 295 K since the relative contribution of highly emissive intrachain excited states is enhanced at a lower temperature. Such hybrid *HJ*-aggregation and the prevalent structural disorder in an IDTBT thin-film bring about a mixed origin of lower-energy emission transitions from both intrachain and interchain excitons, e.g. judged by the clearly enhanced 0-1 vibronic emission peak intensity in the room temperature thin-film spectrum relative to the corresponding solution emission spectrum (Fig. S3).

In addition, the temperature-dependence of red-shifted location of the vibronic emission peaks from 295 K to 140 K allows us to determine fairly small energetic disorder of $\sigma = 8.8$–$12.3$ meV in an IDTBT thin-film (see Fig. S4 in detail), due to long-range uniformity in backbone conformation (9, 32). A slightly larger extent of energetic disorder for the 0-1 emission band indicates that, unlike the 0-0 vibronic optical transition, this red-shifted 0-1 transition originates from both interchain and interchain excitons. Another notable observation from Fig. 1f lies in a close match of the absorption and emission vibronic peak ratios measured at temperatures between 295 K to ~260 K (c.f. Fig. S2), which deviate from the fitting result based on the analytical model for hybrid *HJ*-aggregates (33, 34). We propose to tie these similar spectral peak ratios to the thermal accessibility of all electronic density of states (DOS) at temperatures between 295 K and ~260 K, which then allows an estimate of the energy barrier for thermal access to the tail states of excitonic DOS to be $\approx 260$ K = 22.4 meV in IDTBT thin-film. This value of a thermal activation energy is close to the Urbach energy $E_u = 23$–$24$ meV as a measure of the degree of disorder (or the width) in the joint DOS near the band-edge (30, 35), determined independently from the photothermal deflection spectrum of optical absorption in pristine IDTBT thin films (12, 13, 32).

**Ultrafast Exciton Transport and Nonlinear EEA Dynamics**

The EEA process is a major loss channel at a high exciton density, which therefore should be minimized for energy efficient devices and commercial applications. On the other hand, the competing EEA and excitation transfer dynamics can be used as a probe to investigate exciton transport. A comparison of EEA dynamics in IDTBT solution and IDTBT thin film not only offers the possibility of direct comparison of intra- and inter-chain transport, but can also reveal how a change in microstructure and many-body interaction can influence the exciton-transport process and EEA-induced excitation losses.

Fig. S5 illustrates a series of 'traditional' excitation fluence dependent transient absorption (TA) measurements in the solution and thin film samples, where the pump spectrum of all our TA experiments is resonant with the 0-0 transition, and the detection wavelength is set at the maximum of the ground state bleach (GSB) signal (680 nm). All the transients are normalized to the tail signal averaged between pump-probe delay from 350 to 400 ps. In each sample, an acceleration of GSB recovery is observed with an increase in excitation pulse energy due to the increase in EEA and potentially other two-particle exciton quenching (see below and SI Appendix Section II). The intensity of TA signals recorded at various delay times (see Fig. S6 for details), plotted here as a function of pulse energy, are compared with the corresponding TA intensities recorded from IDTBT thin films at room temperature and 77 K. A linear dependence of the solution TA signal is seen at early time, and a non-linear dependence at longer time due to the occurrence and increasing dominance of nonlinear exciton-exciton annihilation with higher fluences. IDTBT thin films, on the other hand, exhibit a strong nonlinear dependence of the room-temperature TA signal at all plotted delay times when the pulse energy exceeds ~10 nJ ( a 10 nJ pulse corresponds to ~$5\times10^{19}$ photon/cm$^3$) because in this case, IDTBT backbones in the solid state are close packed into an *HJ*-aggregated polymer network, in which the interchain excitonic species and ultrafast EEA kinetics dominate the TA responses. Nevertheless, recovery to a linear dependence on pulse energy was observed in IDTBT thin film at low temperatures of <140 K (e.g. 77 K) since the interchain transport and resulting EEA are significantly suppressed in this temperature range (vide infra).

An excitation fluence dependent TA study is a commonly used method to investigate diffusion-controlled excitation transport in the natural and synthetic light-harvesting systems (36, 37).

However, quantitative isolation of multi-exciton interactions and intra/inter-chain transport from single-exciton relaxation dynamics is a challenge. If the intensity dependent TA signals are dominated by exciton numbers and their dynamics we can apply the phase-cycling based transient absorption method recently proposed by Brixner and co-workers (16), which allows quantitative separation of up to (2N+1)th order nonlinear signals by measuring TA kinetics at *N* prescribed pump-intensities. For example, when the TA signals are measured at three defined pump intensities such as $I_0$ (2 nJ), 3 $I_0$ (6 nJ) and 4 $I_0$ (8 nJ) (shown in the left panel in Fig. 2b for the studied IDTBT thin film), linear combination of the three signals enables isolation of the third (PP3), fifth (PP5) and seventh (PP7) order nonlinear signals,

$$PP3\ I_0 = \frac{1}{4}PP(4I_0) - \frac{2}{3}PP(3I_0) + 2PP(I_0) \qquad \textit{One-particle dynamics} \qquad [1a]$$

$$PP5\ I_0^2 = -\frac{1}{3}PP(4I_0) + \frac{5}{6}PP(3I_0) - \frac{7}{6}PP(I_0) \qquad \textit{Two-particle dynamics} \qquad [1b]$$

$$PP7\ I_0^3 = \frac{1}{12}PP(4I_0) - \frac{1}{6}PP(3I_0) + \frac{1}{6}PP(I_0) \qquad \textit{Three-particle dynamics} \qquad [1c]$$

While the PP3 signal reports single-exciton relaxation dynamics, the PP5 and PP7 signals inform about the EEA dynamics involving up to two- and three-exciton interactions, respectively. In addition, PP5 often exhibits an opposite sign to PP3, as predicted by the response functions, when the transition dipole moments from the single- to two-exciton state is comparable to or smaller than that from the ground to single-exciton state.

**Single-exciton relaxation dynamics**. The isolated room-temperature PP3 kinetic traces in the solution and a thin film of IDTBT copolymer are illustrated in Fig. 2. The extracted PP3 trace in each sample overlaps well with the TA kinetic measured at low pump fluence (for example, at 0.4 nJ shown in Fig. S7) but with an enhanced signal-to-noise ratio, ensuring the successful isolation of higher order nonlinear signals of our analysis. In IDTBT solution, the PP3 kinetic, *i.e.* the single-exciton decay kernel [PP3(*t*)], shows a bi-exponential decay with an average exciton lifetime $\tau_{avg}$=1210 ps. We assign the short time constant (2.1 ps) in Table 1 to hot exciton relaxation, while the longer time constant (1490 ps) is assigned to luminescence decay of the intrachain excitons to the ground state and is similar to the 1.59 ns intensity independent fluorescence decay recorded by time correlated single photon counting. The energetic relaxation of hot excitons occurs mainly through the early on-chain migration and depopulation of excitations, mediated by HOMO-LUMO orbital overlap and the resulting on-chain charge transfer process. In the thin-film, the PP3 kinetic trace exhibits a tri-exponential decay ($\tau_1 = 4.8$ ps, $\tau_2 = 27.6$ ps, and $\tau_3 = 620$ ps; see Fig. S8 for a schematic illustration of the energy levels and decay feature of IDTBT excited states), along with almost a plateau at longer times (>250 ps). The additional intermediate time constant ($\tau_2$), which is not present in solution and likely difficult to detect in conventional TA measurement of IDTBT thin films, is due to the formation of interchain singlet excitons. In general, the optical transition between interchain excitons and the ground state is energetically unfavored and takes place with a rather low emission efficiency in hybrid *HJ*-aggregates; however, we found that the interchain excitons with $\tau_2 = 27.6$ ps are fluorescent (i.e., red-shifted emission centered around the 0-1 vibronic emission peak as a result of the Stokes shift and electronic-vibronic coupling at close chain-chain contacts) based on our thin-film TA measurement probed at 800 nm, as detailed in Fig. S9 which also shows that the photoluminescence quantum efficiency (PLQE) of the emissive interchain excitons for near-infrared (NIR) emission could be four times that of IDTBT intrachain excitons. In addition, strong coupling between an interchain exciton with phonon modes in IDTBT thin-films tends to stabilize a small fraction of long-lived, lower-energy charge transfer/separated interchain excited states with a microsecond lifetime (32) which gives rise to the plateau seen on the PP3 signal within our experimental time window of 700 ps (Fig. S10d). The average exciton lifetime ($\tau_{avg}$=125 ps) of the photogenerated singlet excitons at room temperature is an order of magnitude shorter in thin-film compared that in solution, which we suggest results from the dominant 27.6 ps decay of emissive interchain excitons and the non-emissive nature of a minor phase of the long-lived charge-transfer and/or charge-like interchain excited states.

**Resolving EEA dynamics from higher-order nonlinear responses.** Following Malý et al. (16), the 5$^{\text{th}}$-order nonlinear response for travelling excitons can be written as

$$\text{PP5}(t) = A_0 \left(1 - \exp\left(-\int k_A \, dt\right)\right) \text{PP3}(t) \qquad [2]$$

where PP3(t) still represents the 3$^{\text{rd}}$-order response indicating the single-particle decay dynamics. $k_A$ is the reaction rate for the following annihilation process of singlet exciton pairs ($S_1$), contributing to the observed rise pattern in PP5,

$$S_1 + S_1 \xrightarrow{k_A} S_1 + S_0 \qquad [3]$$

The clear difference shown in Fig. 2c between the room temperature 5$^{th}$-order responses of solution and thin-film samples of IDTBT arises from the variation of two-exciton EEA rise ($k_A$) and one-exciton decay (PP3) terms in Eq. **2**. In the solution, a slow rise of PP5 at longer delays times of >100 ps is clearly observed (Fig. S10e), while in the thin film only a rapid rise peaking at ~6 ps is seen (Fig. 2b). This difference in PP5 rise time is ascribed to the larger value of $k_A$ and the more rapid decay of the PP3 signal for thin film samples, as described by Eq. **2**. The discrepancy between the isolated thin-film PP5 signal and the fits of Eq. **2** may arise from the single-particle dynamics associated with long-lived interchain charge-transfer states, which result in the long tail on the PP3 traces but significantly weaker two-particle transient response (see details for extended two-component two particle decay analysis in SI Appendix Section II).

To obtain physical insight from the $k_A$ values, we approximate the migration behavior of excitons as a diffusive process on IDTBT chains. If the annihilation kinetics is limited by exciton migration and an encounter rate, we can describe the annihilation process using a diffusion-limited kinetic model. Furthermore, analysis of the 5$^{th}$ order nonlinear transient signals with a diffusion model requires consideration of the dimensionality of exciton motion. We applied one- (1D) and three- (3D) dimensional kinetic models to analyze the isolated PP5 signals (modelling details provided in Methods Section III and Fig. S11) and show our fitting results in the right panels of Fig. 2. According to these model fits, the exciton transport can be described as 1D in solution, while it is 3D in an IDTBT thin film due to the formation of interchain interactions at a high density of close chain-chain contact sites. Interchain interactions at short chain-chain contacts allow excitons to hop from one chain to an adjacent chain, in contrast to the isolated IDTBT chains in solution where only 1D on-chain transport is possible.

A potential concern in the thin film analysis is the possible presence of exciton quenching by long lived charge-like states, which is also a two-particle phenomenon (38). Thomas et al. (39) report spectrally resolved transient absorption spectra for IDTBT among several polymer species. They find evidence for long-lived interchain excited states, but with a rather small population in IDTBT relative to the total number of interchain excited states. These long-lived charge transfer/separated states likely give rise to the long tail on our PP3 ground state recovery signal, while still contributing to the two-particle PP5 response. Here, we show that the PP3-extracted decay time in IDTBT thin film agrees well with its fluorescence lifetime of ~130 ps (Methods Section I); this indicates the relative fraction of such non-emissive charge-like states beyond our probe window and the associated occurrence of additional non-radiative decays should be small in the thin film samples, which also remains consistent with our photocurrent results demonstrating these non-emissive long-lived interchain excited states of IDTBT are localized (not very mobile) in the thin film and their relative population is not significant, relative to the total number of excitonic states detected by the phase-cycled TA spectroscopic measurements (In SI Appendix Section III we describe two component fits of PP5 with varying amplitudes of charged species, and show that the resulting $L_d$ values differ rather little from that obtained from single component fit of PP5.).

In Table 1, we report the exciton migration diffusivity ($\mathcal{D}$) and diffusion length ($L_d$) at room temperature calculated by Eq. 2 and the diffusion models, based on the 5$^{th}$-order data which give $\mathcal{D} = 0.56$ cm$^2$/s (and $L_d = 367$ nm) and $\mathcal{D} = 0.07$ cm$^2$/s (and $L_d = 75$ nm) in IDTBT solution and thin-film, respectively. A room-temperature exciton transport mobility of $\mu = 2.9$ cm$^2$ V$^{-1}$ s$^{-1}$ in IDTBT thin-film can be obtained from Einstein's relation ($\mu = \mathcal{D}/k_B T$) using its PP5-fitted diffusivity. Analogously, the calculated 1D diffusivity for IDTBT solution gives $\mu \approx 2000$ cm$^2$ V$^{-1}$ s$^{-1}$ along the isolated IDTBT backbone. The $L_d$ in the thin film is ~5 times shorter than that in the solution phase, and this value can be an underestimate due to the assumption of diffusion-limited kinetics in our kinetic model. Based on our PP7 analysis described below, the EEA is an annihilation-limited process in thin-film phase, and the extracted value of $k_A$ ($k_{fit}$ in Table 1) is dominated by the annihilation probability (vide infra). Therefore, the actual exciton encounter rate (and exciton diffusivity) may be even higher than the value that we calculated from a diffusion-limited model. Our results show that both diffusivity and diffusion length values for IDTBT thin film are among the highest for large-area-extended continuous semiconducting polymer films reported to date (c.f. Fig. 3a for comparison of the diffusion constants of various conjugated polymers and representative molecular semiconductors), and are consistent with a highly rigid IDTBT backbone and a low degree of energy disorder in IDTBT thin films.

**7$^{th}$ Order Signals and Reaction Regime of EEA Dynamics**

To further explore the effects of many-body interactions on the exciton motion and higher-order EEA dynamics, we turn to the 7$^{th}$ order pump-probe signals (PP7). The 7$^{th}$ order nonlinear response signal involves 7 light-matter interactions and contains the response function for up to 3-exciton dynamics. These multi-exciton interactions yield insights into the 'annihilation probability ($P_A$)' when multi-excitons encounter each other. The response function for the 7$^{th}$ order signal (PP7), describing dynamics involving up to three excitons in 3D, is given by:

$$PP7(t) = A \times PP3(t) \times \left\{1 - \frac{2f-3}{f-1}e^{-k_A t} + \frac{f-2}{f-1}e^{-fk_A t}\right\} \quad [4]$$

where A is the response amplitude; $PP3(t)$, and $1 - \frac{2f-3}{f-1}e^{-k_A t}$ represents the single and two exciton dynamics, respectively. In the last term, ($fk_A$) represents the three-exciton annihilation rates where $k_A$ is the two-exciton dynamics in the PP5 response. Here, the *f* parameter provides the relationship between the two- and three-exciton annihilation rates, determined by the following cases of different annihilation probability:

(*i*) If the annihilation probability is very low ($P_A \ll 100\%$), colliding excitons are allowed to pass through each other with a probability of $1 - P_A$ and then continue to travel freely to encounter other excitons, with the $k_A$ being limited by the annihilation probability. In this scenario annihilation events will occur independently between different pairs of excitons, with the case for three-excitons expressed as $\binom{3}{2} = 3$ combinations of exciton pairs selected from three interacting excitons. In this annihilation-limited case, the *f* value equals 3 due to three combinations of available exciton pairs compared to the two-exciton interactions.

(*ii*) On the other hand, if the annihilation occurs with a near-unit probability ($P_A \approx 100\%$) which resembles the classical EEA kinetic regime, excitons will interact strongly upon first encountering each other, and the diffusive trajectory of the surviving excitons becomes the dominant factor determining $k_A$. In this diffusion-limited case, the *f* value is not solely related to the increased combination of exciton pairs, which accelerates $k_A$ by 3-fold; rather, the exciton encounter rate is also accelerated from the three-exciton annihilation rate $fk_A$, as the first-encounter time is shortened simultaneously in the presence of three-exciton interactions as compared to two excitons. The *f* value in the diffusion-limited case eventually becomes larger than 3 (*f* > 3) due to the acceleration from these factors.

(*iii*) In cases where the three-exciton dynamics are absent, the last term in Eq. **4** is eliminated with *f* = 2, thereby returning the same form of Eq. **2** for only pair-wise EEA dynamics.

As a result, differentiating the patterns of PP7 signals provides a simple method to experimentally distinguish between the annihilation-limited ($f = 3$) and the diffusion-limited ($f > 3$) regime. As Fig. S13 shows the last term $1 - \frac{2f-3}{f-1}e^{-k_A t} + \frac{f-2}{f-1}e^{-fk_A t}$ in Eq. **4**; when $f = 3$ PP7 shows a monotonously increasing trend, whereas for cases with $f > 3$ PP7 exhibits a sharp decay followed by a steady increase to positive values. Accordingly, Fig. 4 indicates that $f \approx 3$ for IDTBT in thin-film form at all measured temperatures, while $f > 3$ for IDTBT in solution. We will return to annihilation limited result when we discuss the temperature dependences of the overall EEA rate and the macroscopic photocurrent generation in IDTBT thin films. The $f$ parameter obtained from fits of the thin-film 7$^{th}$ order data, however, remains constant at $3 \pm 0.4$ over the whole temperature range from 295 K down to 77 K. This suggests that the annihilation-limited regime relies on the microscopic chain-chain contact pattern and weak interchain coupling in IDTBT thin films, rather than quantum interference induced suppression of EEA proposed to arise from stronger interchain interactions in *H*-aggregated molecular micro-crystals at cryogenic temperatures (31).

**Temperature Dependence of $k_A$ and Photocurrent in IDTBT Thin-film**

In order to explore how temperature affects the competition between exciton transport and EEA dynamics, a set of phase-cycled TA measurements were carried out over a temperature range of 295 K to 77 K (c.f. Fig. S14 for detail). The single-exciton average lifetimes ($\tau_{avg}$) and two-exciton annihilation rate ($k_A$) were extracted. In the IDTBT thin film, decrease in temperature increases the $\tau_{avg}$, and the EEA rate decreases (see also Table S1). In contrast, both $\tau_{avg}$ and the EEA rate in a 50 wt% PCBM blended IDTBT thin film as control sample (data not shown here) are almost temperature independent, likely due to the localized charge-transfer state or attenuation of the chain-chain interactions of IDTBT by the surrounding PCBM molecular acceptors. This difference suggests that the exciton interchain migration in IDTBT thin films require thermal activation whereas the on-chain transport does not.

To complement our TA results, we recorded the temperature-dependent photocurrent response in IDTBT thin-film transport channels (Methods Section IV) and compared it with the single-exciton relaxation and two-exciton EEA dynamics. The exciton interchain hopping is a limiting step for the amplitude of the photocurrent signal as the critical in-plane dimension of a polymer thin-film transport channel (here, channel length = 12.5 µm shown in Fig. S15) far exceeds the length of the polymer backbone (*i.e.* 426 nm estimated based on the weight average molecular weight, $M_w$ = 346 kg mol$^{-1}$ of the IDTBT copolymer used). Here, the photocurrent signals recorded across an IDTBT film channel require the photo-generated excitons to move close to the gold electrodes and then dissociate into free charges to result in a photocurrent signal. The photocurrent signal in Fig. 5a displays a clear modulation by about a factor of ten in the current flow through an IDTBT thin-film transport channel when 656.1 nm illumination on the channel was turned on versus off. Both the on-off photocurrent difference and the PP5-extracted exciton diffusion length decrease with decrease in temperature (Fig. 5b & Fig. S16).

Furthermore, from Fig. 5c it is evident that both the photocurrent and the EEA rate follow the same temperature dependence at high temperatures (>170 K), indicating that both exciton interchain hopping and two-exciton annihilation frequency are thermally activated. In Fig. 5b, an Arrhenius fit of the temperature-dependent photocurrent data for T ≥140 K gives a thermal activation energy barrier for interchain hopping of $E_a$ = 57.4 ± 2.5 meV, which is among the lowest for device-relevant continuous organic semiconductor films (Fig. 3b). Below 140 K, the photocurrent signal in an IDTBT thin-film channel is much smaller since interchain hopping is strongly suppressed in this low temperature range. This suggests that a structural reconfiguration is required for a successful exciton interchain hopping event, the thermal energy at lower temperatures is no longer able to surmount the potential energy barrier associated with structural rearrangement and/or energy disorder.

A transition at 140 K is also seen for the average exciton lifetime as a function of inverse temperature (Fig. 5b), with the slope of the fits above and below 140 K differing by a factor of 3.2. This is consistent with strong suppression of interchain emission and transport in IDTBT thin film at temperatures below 140 K. Moreover, the changing trends of the photocurrent signal and the PP5-extracted annihilation rates (and $L_d$) start to deviate at low temperatures as well (see Fig. 5c and Fig. S16), in line with the progressively suppressed interchain transport whilst the survival of on-chain transport and intrachain exciton annihilation at temperatures below 140 K.

## Discussion

The formation of interchain excitons (and long-lived charge transfer products) in semiconducting polymer films requires structural and interaction stabilization of adjacent chains at close chain-chain contact sites. The generation efficiency of the emissive interchain excitons in IDTBT thin films was shown not to be dependent on pump fluence, but determined by the density of close chain-chain contacts formed in IDTBT thin film (39). We could not resolve this emissive interchain excitonic species through time-correlated single photon counting (TCSPC) measurement due to the short lifetime of 27.6 ps. We showed that these fluorescent interchain excitons emit NIR light more efficiently than do the intrachain excitons (Fig. S9). Fig. 5b further demonstrates the thermal activation character of the overall thin-film lifetime and thus the fraction of emissive interchain excitons, given the lifetime of intrachain excitons in IDTBT thin films remains nearly constant based on the isolated PP3 profiles at different temperatures.

When there is a substantial density of close chain-chain contacts in an IDTBT thin film, these emissive interchain excitons dominate the photogenerated excited states (rather than the intrachain excitons). Also, the size and mobility of such interchain excitons can be sufficient to govern EEA kinetics due to the planar conformation of IDTBT backbone. These observations are consistent with the conclusion that the EEA rate in IDTBT thin films is largely dominated by the kinetics of emissive interchain excitons (see also Fig. S17 and Fig.S18). In this way, the short lifetime of 27.6 ps can also be regarded as a direct result of much faster two-particle EEA rates between interchain excitons than that of intrachain excitons (see Fig. S17 and SI Appendix Section II for details). Whether an interchain excited state is fully or partially charge-separated strongly affects its lifetime. Considering the emissive feature and short lifetime, we expect the emissive interchain excitons ($\tau_2$=27.6 ps) to lack strong charge-separation character, and their NIR-fluorescent decay to the ground state most likely occurs when they are spatially pinned or aggregated by interchain interactions at a close chain-chain contact site (4, 39).

The two-exciton EEA rate, $k_A$, between 300 K and 77 K in the thin film of the high-mobility IDTBT copolymer shows a significant temperature dependence with a substantial increase, with data above 140 K fitting well to the Arrhenius form with a fairly small activation energy (Fig. 5b). The amplitude of the photocurrent flow in the IDTBT thin-film is also strongly temperature dependent with a very similar dependence to that of the $k_A$ for the thin film (refer to Fig. 3c); in both cases the activation energy is 57.4 meV (460 cm$^{-1}$). An activation energy for photocurrent generation is not surprising as a Coulomb barrier must be surmounted to separate the charges (40, 41). The fact that IDTBT side chains are non-polar should make this barrier small. Venkateshvaran et al. (12) find the energetic disorder in IDTBT thin films to be 23 meV (194 cm$^{-1}$). The additional 33.4 meV involved in photocurrent generation ($E_a$ = 57.4 meV) may represent the Coulomb barrier contribution. If this is the case, it may indicate that the EEA process and excitonic interactions in IDTBT copolymer also involves a charge transfer character followed by rapid energy and structural relaxation leading to formation of the ground state. The activation energy for EEA could also, then, be the underlying reason for the annihilation probability being significantly less than unity in IDTBT thin films.

A recent modeling study by Makki et al. (42) suggests that the BT-BT π-π interactions with a preference for perpendicular orientations provide the pathway for the interchain charge transport in the solid state of IDTBT copolymer. The calculated density of the chain-chain crossing points mediated by polymer chain self-assembly are substantial, while the calculated interchain electronic couplings range from 0.01 eV up to 0.1 eV, which are also in the range of the Stokes shift and energy disorder identified in the pristine IDTBT thin-film from this work. As noted above the activation energy governing the exciton interchain hopping and the two-exciton EEA dynamics may arise from a structural reorganization that inhibits recombination of travelling excitons (43–45). This reorganization via electron-phonon coupling may also lie at the root of the annihilation-limited EEA dynamics seen in the film, along with the temperature dependence of $k_A$. One finding suggesting that not all interchain contacts can be perpendicular BT-BT interactions lies in the mixed *HJ*-like aspects of the optical absorption and emission spectra. These imply that there must be stretches of parallel interchain interactions resulting in the hybrid *J*- and *H*-aggregate spectra of thin film samples. This is consistent with high-resolution imaging results of IDTBT films and interacting chains, which visualize local regions of IDTBT copolymer with the backbones aligned in parallel and with perpendicular crossing as well as non-ordered regions (46, 47).

Strikingly, we demonstrate that the nearly amorphous IDTBT thin films exhibit a large diffusion coefficient despite the short exciton lifetime: an exciton diffusion length of $L_d$ = 75 nm within an exciton lifetime of 125 ps, corresponding to a supersonic velocity of 600 m/s. This is enabled by the annihilation-limited transport regime, and the low site-energy fluctuations together with the ultra-long persistence length measured for IDTBT chains (48). In an IDTBT thin-film, multiple exciton encounters and separations over the exciton lifetime enhances the average distance of hopping trajectories. A narrow distribution (in accord with the small energy disorder) of the excitonic DOS should enable a large migration mobility along the planar backbones and also a high density of available sites for subsequent interchain hops.

## Concluding Remarks

Our analysis of the 5th order pump-probe signals show that the dimensionality of the exciton motion is very different in IDTBT solution from in the thin film. The solution-phase data fit best to a 1D model in which the EEA kinetics are diffusion limited. In contrast, the thin film 5th order signal, in combination with the 7th order response, fits to a 3D model in which the EEA probability is <<100% per encounter. Thus, excitation losses resulting from EEA at high excitation densities will be significantly reduced in the thin films compared to the diffusion-limited case. The large velocities (600 m/s thin film, 303 m/s solution) imply that simple Förster-type hopping transport is not an adequate description of the exciton motion. The thermal accessibility of the electronic density of states may enable the type of large steps proposed in Ref. (49) and clearly investigating the mechanism of exciton motion will be very worthwhile in this remarkable copolymer. Multi-wavelength transient absorption studies along with mixed ultrafast vibrational and electronic spectroscopies would be valuable additions to the studies reported here.

Both solution-phase and thin film forms of IDTBT exhibit large excitation diffusion lengths. In the solution case the $L_D$ of 367 nm is an unusually high fraction of the total chain length (estimated to be 426 nm). In IDTBT thin film $L_d$ = 75 nm which is twice the thickness of the thin film. In general, device fabrication and performance benefit from large diffusion lengths and annihilation limited transport with a rather low EEA probability per encounter. In particular, if the diffusion length of excitations exceeds the thickness of an organic semiconducting layer, bulk-heterojunction fabrication may not be necessary, potentially allowing for building a simple single-layer architecture of organic photovoltaic devices. When $L_d$ is larger than the critical dimension of catalytic or sensing nanostructures, it can be expected to mitigate or prevent the formation of so-called dead volume where the generated excitations are trapped inside within lifetime and therefore hardly arrive at the outer active sites; in this way, the reaction activity and overall power efficiency can be upgraded simultaneously.

## Methods

**I. Sample preparation, Steady-state optical spectra and Fluorescence lifetime**

IDTBT copolymer was synthesized via a Suzuki cross-coupling condensation (50). IDTBT solution samples were prepared by dissolving IDTBT powder ($M_w$ = 346 000, polydispersity index, PDI = 2.43) in 2-methyltetrahydrofuran (≥99.5%) to form a clear 0.15 mg/ml solution, which was transfer into a quartz cuvette for measurements. IDTBT thin films were spin-coated on pre-cleaned polished Spectrosil substrates from a precursor solution (10 mg/ml IDTBT in 1,2-dichlorobenzene). All solution processing took place under a $N_2$ atmosphere. The solution and film samples were then accommodated in a cryostat to record temperature dependent steady-state absorption and photoluminescence spectra. Time-correlated single photon counting (TCSPC) fluorescence lifetime measurements of IDTBT solution and thin-film samples were excited at 404 nm and detected at 710 nm, but only the solution sample was found to give strong enough photon signals for a reliable extraction of fluorescence lifetime as ~1.6 ns (see Fig. S1). This is because the IDTBT film has a low photoluminescence quantum efficiency (PLQE) of 1.7% in relation to the PLQE value of 21% determined from IDTBT solution using an integrating sphere method (39, 51). This small PLQE value gives a thin-film fluorescence lifetime of ~130 ps compared to the 1.6 ns that we have determined for IDTBT solution, which is in excellent agreement with the average thin-film singlet lifetime ($\tau_{avg}$ = 125 ps) in Table 1.

**II. Phase-cycling-based TA spectroscopic measurement**

A Ti/sapphire regenerative amplifier (Coherent, RegA 9050) seeded by a Ti/sapphire oscillator (Coherent, MIRA Seed) and a diode-pumped laser (Coherent, Verdi V-18) was used to generate 800 nm pulses with a repetition rate of 250 kHz. The 800 nm beam was split using a beam splitter. To generate the probe, the beam was focused on a 1 mm sapphire crystal to produce a visible continuum. Following continuum generation, a 700-nm short-pass filter was placed to filter out undesired wavelengths. To generate the pump, an optical parametric amplifier (Coherent, OPA 9450) was tuned to generate pulses centered at 680 nm resonant to the 0-0 transition of IDTBT. The pump pulses had a FWHM of 70 fs. The pump and probe were 160 μm and 80 μm in diameter, respectively, and the beams were overlapped at the sample at the magic-angle (54.7°) polarization. The systematic error of pump fluence due to the Gaussian beam profile, comparing the average fluence to the edge, is approximately 20% (see detailed comparison in SI Appendix Section V). The probe was passed through a monochromator (Acton Research Corp., SpectraPro 300i). The signal was detected using a Si-biased photodiode (Thorlabs, DET10A) connected to a lock-in amplifier (Stanford Research, SR830), which was synchronized to a chopper positioned in the pump beam path. Each set of phase-cycling based pump-probe transient was collected at three different pump pulse energies: 2, 6 and 8 nJ for the IDTBT thin film, and 6, 18 and 24 nJ for the IDTBT solution. For temperature dependent measurement we used liquid nitrogen cooled cryostat from Oxford instrument (OptistatDN) to control the temperature.

### III. Diffusion model for 5$^{th}$-order nonlinear response

In the data analysis, the fitting equations for 5$^{th}$ order nonlinear signal are derived from the excitonic manifolds model from the work of Malý et al. (16). The population evolution of excitonic manifolds is written as

$$d\mathbf{P}/dt = -\mathbf{K}\,\mathbf{P} \qquad [5]$$

where the vector $\mathbf{P}$ is the population of each excitonic manifold, and $\mathbf{K}$ is the rate matrix of possible decay pathways, including excitation relaxation and annihilation for 2-exciton manifold. The 5$^{th}$ order nonlinear response function is obtained by summing all possible responses in double-sided Feynman diagram (16), and written as:

$$PP5(t) = A\big(1 - exp(-\textstyle\int k_A\,dt)\big)\,PP3(t) \qquad [6]$$

where $k_A$ is the 2-exciton annihilation rate described in Eq. **2**. In the modeling, the exciton-exciton annihilation processes are approximated as a diffusion-limited reaction, enabling the substitution of diffusion reaction rate ($k$) for the 2-exciton annihilation rate ($k_A$). The relationship between diffusion rate and diffusion constant ($D$) varies with the dimensionality of exciton transport according to

$$(1D)\ k = \sqrt{\frac{8D}{\pi t}} \qquad [7]$$

$$(3D)\ k = 4\pi R D \qquad [8]$$

where $R$ is reaction radius of annihilation. By substituting $k_A$ in Eq. **6** with the diffusion rates of Eq. **7** or **8**, we obtain the derived PP5 equation for the 1D or 3D system:

$$(1D)\ PP5(t) = A\left(1 - exp\big(-\sqrt{32Dt/\pi}\big)\right) PP3(t) \qquad [9]$$

$$(3D)\ PP5(t) = A\big(1 - exp(-4\pi R D t)\big)\,PP3(t) \qquad [10]$$

During the fitting process, the exponents in Eq. **9** & **10** are simplified by $k_{fit} = 4\pi RD$ or $k_{fit} = \sqrt{32D/\pi}$, respectively, where $k_{fit}$ is a fitting parameter. For 3D systems, $k_{fit}$ equals the two-exciton annihilation rate $k_A$ and is directly reported as $k_A$ in the main text.

To convert $k_A$ into exciton diffusion constant, we assume: (*i*) the excitons travel in a chain with the weight-averaged IDTBT backbone length of 426 nm and 1.8 nm width, or in 3D with 0.41 nm π-π stacking distance along the thickness direction of a thin film (50, 52). (*ii*) an EEA event occurs immediately when two excitons encounter each other within a 1.6 nm reaction radius corresponding to the length of the repeating IDTBT monomer (42, 46). The diffusion length ($L_d$) of excitons is then calculated from the diffusion constant by $L_d = \sqrt{2ND\tau}$, where N is dimensionality and $\tau$ is the average exciton lifetime from the isolated PP3 signals.

The annihilation rate ($k_A$) involves the exciton encounter rate (controlled by exciton migration), annihilation probability per encounter of excitons, and annihilation rate for the 2-exciton manifold ($k$). According to the PP7 analysis (main text), the EEA dynamics in IDTBT thin films are in an annihilation-limited regime. In this case, the annihilation probability is the rate-determining factor of the $k_A$, leading to a breakdown of the substitutability between $k_A$ and diffusion constant $k$. As a result, applying the diffusion-limited model in an annihilation-limited regime may underestimate the diffusion constant for IDTBT thin films since $k_A$ is reduced by the annihilation probability.

**IV. Temperature-dependent measurement of photocurrent generation**

IDTBT thin-film transport channels (channel length = 12.5 µm & width = 75 µm, and with deposition of no gate electrode) were made from the IDTBT thin-films spin-coated using the same recipe for fabricating the TA thin-film samples (c.f. Fig. S15). Square source and drain contact electrodes of 50 nm-thick Au were then deposited on top of an IDTBT thin film, via electron beam evaporation using metal mask under high vacuum. Temperature-dependent photocurrent response measurements were performed in a cryogenic probe station (LakeShore) with a 4155C Semiconductor Parameter Analyzer (Agilent Technologies), when the illumination of a laser diode (λ = 656.1 nm, power = 4.4 mW) on the transport channel area was kept on or switched off. The laser beam size on the thin-film samples is 2 × 4 mm rectangular. All on-off sequence of the photocurrent signals at 295 K and 77 K were implemented with source-drain voltage of 10 V.

**Acknowledgements**

This research was supported by the US Department of Energy, Office of Science, Basic Energy Sciences, Chemical Sciences, Geosciences, and Biosciences Division. Y.S. and K.L. acknowledge additional financial support from the Max Planck Society. N.H. acknowledges support from JST PRESTO (JPMJPR23H7), Japan. A.J. acknowledges support from the U.S. Department of Energy, Office of Science, Office of Basic Energy Sciences, Materials Sciences and Engineering Division under contract no. DE-AC02-05Ch11231 (Electronic Materials program). We are grateful to Paul Blom at Max Planck Institute for Polymer Research for helpful discussion. All authors thank Naomi Ginsberg, Omar Yaghi, and Peidong Yang at University of California, Berkeley for allowing facility access.

**Contributions**

Y.S. and G.R.F. proposed the project. Y.S. fabricated samples. Y.S. and P.P.R. measured steady-state optical spectra and TCSPC lifetime. Y.S., P.P.R. and T.-Y.L. recorded phase-cycling TA measurement and analyzed the separated single-, two- and multiple-exciton dynamics. Y.S., N.H. and Q.L. carried out photocurrent measurement. I.M. synthesized IDTBT copolymer. Y.S., G.R.F., P.P.R. and T.-Y.L. drafted the manuscript and all authors contributed to revise the manuscript.

**Competing Interests**

The authors declare no competing interests.

**Data availability**

The authors declare that all study data are included in the article and/or supporting information. The data presented in this study are available from the corresponding authors upon reasonable request.

**Figures and Table**

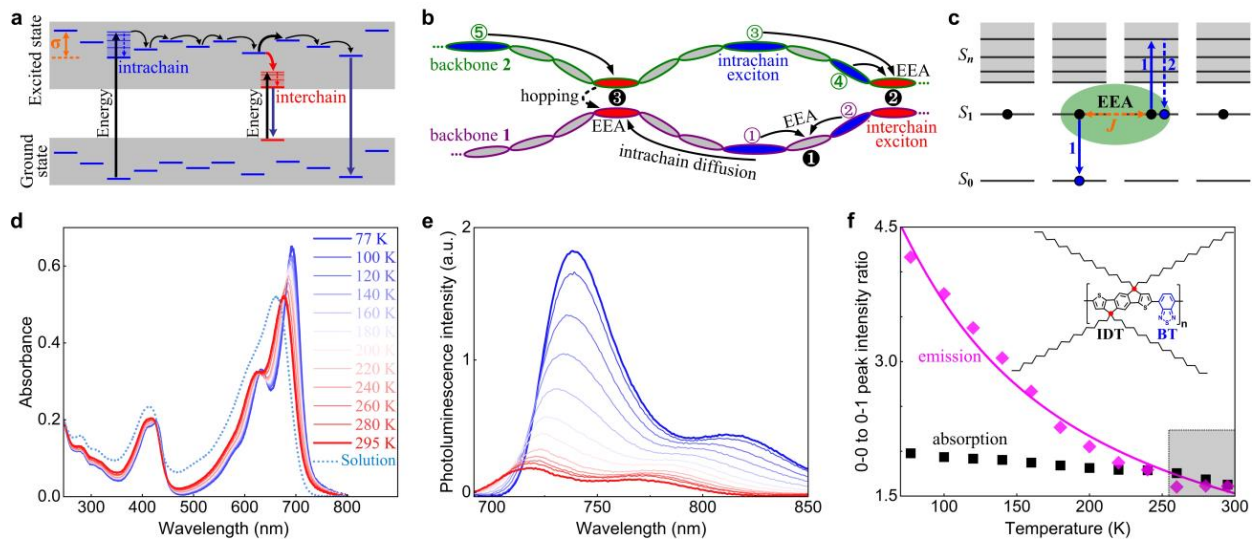

**Fig. 1. (a−b)** Schematic of the competing transport and relaxation pathways of a distribution of excited intrachain (blue lines or ellipses) and interchain (red lines or ellipses) singlet states, and the classical diffusion-limited transport **(b)** in which the exciton relaxation is shown to be terminated by exciton-exciton annihilation (EEA) events occurring with unit probability between two or more excitons, such as the intrachain-exciton pairs marked by ①and ②, ③and ④, ①and ⑤, colliding at an intrachain site (❶) or an interchain short-contact site (❷ and ❸). **(c)** Schematic representing the classical EEA mechanism between two closely approaching (within an annihilation radius depicted with the green-shaded ellipse) singlet excitons. Overall, an EEA event lowers a singlet exciton ($S_1$) to the ground state ($S_0$), while promoting the other interacting singlet to a higher-lying excited state ($S_n$) that subsequently relaxes back to $S_1$: $S_1 + S_1 \xrightarrow{1} S_n + S_0 \xrightarrow{2} S_1 + S_0$. **(d−f).** Temperature-dependent steady-state optical absorption (**d**) and emission (**e**) spectra of IDTBT thin-film, and ratios between the 0-0 and 0-1 vibronic peak intensity in the emission and absorption spectra (**f**). The same color coding was used in (**d**) and (**e**) for the measured temperatures. Also shown is the peak-normalized room-temperature absorption spectrum of IDTBT solution (dotted line) in (**d**) and the best-fit (solid curve) of a $T^{-1}\exp(-\Delta/k_BT)$ dependence for the 0-0 to 0-1 emission vibronic peak ratios in (**f**). The grey-shaded square on the bottom right corner of (**f**) highlights a temperature range in which the emission vibronic ratio remains almost constant; the inset of (**f**) shows the chemical structure of the donor-acceptor monomeric unit of IDTBT copolymer, where IDT donor and BT acceptor moieties are highlighted in black and blue, and the two sp$^3$ bridging carbon atoms in the ring-fused IDT moiety are marked with red dots.

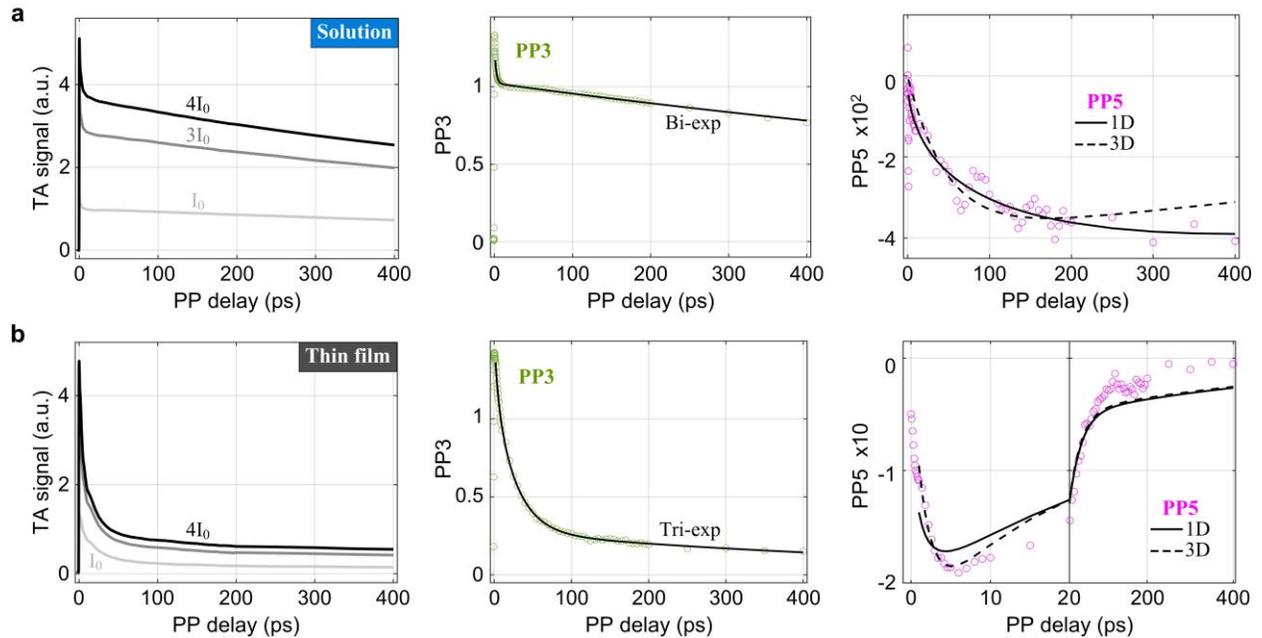

**Fig. 2.** Single- and two-exciton dynamics from the phase-cycled TA spectroscopy in **(a)** IDTBT solution and **(b)** IDTBT thin-film. Left panel: the kinetic TA profiles measured at three prescribed pump pulse energies: $I_0$ = 2 nJ, 3 $I_0$ = 6 nJ and 4 $I_0$ = 8 nJ for thin film; $I_0$ = 6 nJ, 3 $I_0$ = 18 nJ and 4 $I_0$ = 24 nJ for solution. Here, a 6 nJ pulse of pump laser generates 4.6 excitons per IDTBT chain on average, which confirms that we were measuring in a multi-particle interaction regime. Middle panel: isolated third-order (PP3) nonlinear kinetic signal (green circle) and fits (solid black line) using bi- and tri-exponential PP3($t$) decay kernel for the solution and thin-film, respectively. The initial 0.5 ps of all transients was excluded during fitting to avoid possible artifacts caused by the pump-probe pulse temporal overlap. Right panel: isolated fifth-order nonlinear signal (PP5, magenta circle) and fits with 1D (solid black line) and 3D (dotted black line) diffusion model. For the thin-film PP5 kinetic trace, the initial 20 ps is expanded to demonstrate the rapid rise as a result of the ultrafast EEA dynamics.

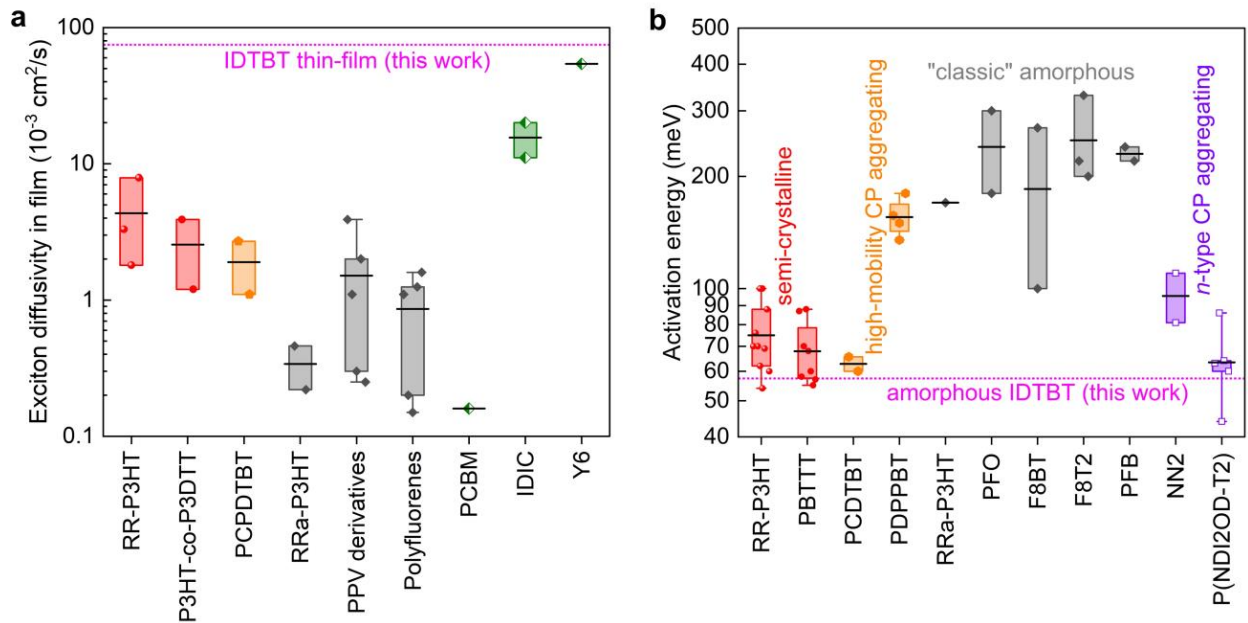

**Fig. 3.** (**a**) 3D exciton diffusivity and (**b**) activation energy of interchain or intermolecular hopping reported in device-relevant continuous films of various representative semiconducting polymers, in comparison to the corresponding results obtained for the amorphous IDTBT thin-films from this work based on the PP5 signal and temperature-dependent photocurrent response. All data are presented in a Logarithm scale and are binned by conjugated polymer (CP) materials for traditional semi-crystalline materials (red), relatively new high-mobility aggregated copolymers (orange), and classic amorphous polymers (grey). Also included are some other types of well-studied high performers, namely molecular acceptor films (green) in (**a**) and aggregated *n*-type polymer films (purple) in (**b**). Data for each material are box-plotted (percent 25th, 75th) with an average (black line). For the complete list of materials and references see SI Appendix Section IV.

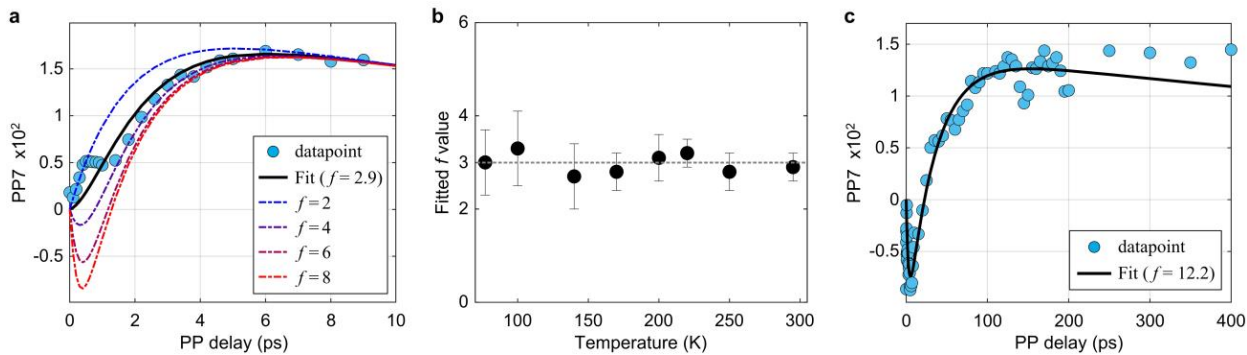

**Fig. 4.** (**a**) Isolated room-temperature PP7 kinetic trace (filled blue circles) in IDTBT thin-film and corresponding spectral fit (black line) using 3D exciton diffusion model, which gave rise to an estimate of 2.9 for the $f$ parameter. Also included in (**a**) for direct comparison is simulated PP7 kinetics (dashed curves) for IDTBT thin-film, based on various $f$ values while keeping all other parameters the same as those obtained from simultaneous fits to the PP3 and PP5 signals. (**b**) Temperature-independence of the fitted $f$ values in IDTBT thin-film; the horizontal dashed line delineates $f = 3$. (**c**) Room-temperature PP7 traces in IDTBT solution, alongside a solid curve of best fit using a 1D diffusion model giving a value of the $f$ parameter of 12.2.

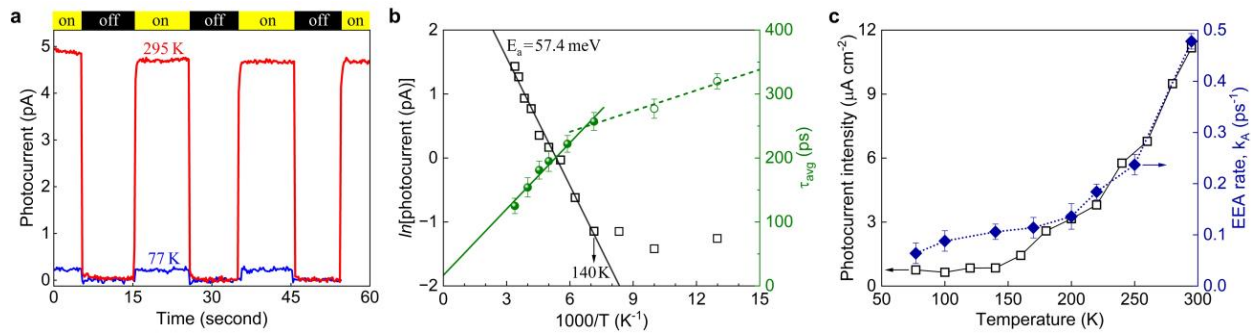

**Fig. 5.** Correlation of temperature-dependent exciton transport and EEA dynamics with photocurrent response in IDTBT thin-film. **(a)** Time sequence of photocurrent signal acquired with source-to-drain voltage of 10 V at 295 K and 77 K when the illumination on the IDTBT thin-film transport channel was turned on and off. The photocurrent decay during the first illumination period may be ascribed to the effect of electrical filling of the thermally accessible tail states of the DOS. **(b)** Inverse temperature ($1/T$)-dependence of both the photocurrent data (in natural logarithm scale, black box) and the average exciton lifetime (calculated from the PP3 fitted time constants). The solid black line corresponds to an Arrhenius fit of the photocurrent data from 295 K to 140 K, which estimates the energy barrier of thermally-activated interchain hopping to be $E_a = 57.4$ meV. **(c)** Comparison of temperature-dependent photocurrent intensity (open box) and two-exciton EEA rate (filled diamond) in IDTBT thin-film.

**Table 1**. The parameters obtained by fitting the isolated PP3 and PP5 kinetic traces at room temperature in IDTBT solution and thin film. The average lifetimes ($\tau_{avg}$) are calculated from the amplitude weighted average of the fitted time constants. $k_{fit}$ is defined in Methods Section III and equivalent to $k_A$ for 3D diffusion model.

| Sample | PP3 kinetics | | | | PP5 kinetics | | |
| --- | --- | --- | --- | --- | --- | --- | --- |
| | $\tau_1$ (ps) | $\tau_2$ (ps) | $\tau_3$ (ps) | $\tau_{avg}$ (ps) | $k_{fit}$ | Diffusivity ($10^{-2}$ $cm^2$ $s^{-1}$) | Diffusion length (nm) |
| Solution | 2.1 (±0.2) | - | 1490 (±30) | 1210 | 0.056 ps$^{-0.5}$ [1D] | 55.87 | 367 (±17) |
| Thin Film | 4.8 (±0.5) | 27.6 (±2) | 620 (±80) | 125 | 0.478 ps$^{-1}$ [3D] | 7.48 | 75 (±3) |